\documentclass[nonacm=true,sigconf]{acmart}
\renewcommand\footnotetextcopyrightpermission[1]{} 
\usepackage{tikz}
\usepackage{mathtools, cuted}
\usepackage{multicol}
\usepackage{svg}
\usepackage{comment}
\usepackage{xcolor}
\usepackage{multirow}
\usepackage{enumitem}
\usepackage{fontawesome}
\usepackage{graphicx}
\usepackage{supertabular}
\usepackage{caption}
\usepackage{subcaption}
\usepackage{amsmath}
\usepackage{amsthm}
\usepackage{bm}
\usepackage{lineno}
\usepackage{tikz}
\usepackage{mfirstuc}
\usetikzlibrary{positioning, fit, calc}   
\tikzset{block/.style={draw, thick, text width=2cm ,minimum height=1.3cm, align=center},   
line/.style={-latex}     
}

\newcommand{\repourl}{https://github.com/abdurrezzak/sBERT-Rank}

\makeatletter
  \if@ACM@anonymous
    \renewcommand{\repourl}{https://anonymous.4open.science/r/sBERT-Rank-7CE6}
  \fi
\makeatother

\setcopyright{rightsretained}
\usepackage{breqn}

\begin{document}

\title{Keywords for Bias}

\author{Abdurrezak Efe}
\affiliation{%
  \institution{Huawei Turkey R\&D Center}
  \city{Istanbul}
  \country{Turkey}}
\email{refe.fb@gmail.com}

\author{Gizem Gezici}
\affiliation{%
  \institution{Huawei Turkey R\&D Center}
  \city{Istanbul}
  \country{Turkey}}
\email{gizem.gezici@huawei.com}

\author{Aysenur Uzun}
\affiliation{%
  \institution{Huawei Turkey R\&D Center}
  \city{Istanbul}
  \country{Turkey}}
\email{aysenur.uzun1@huawei.com}

\author{Uygar Kurt}
\affiliation{%
  \institution{Huawei Turkey R\&D Center}
  \city{Istanbul}
  \country{Turkey}}
\email{uygar.kurt1@huawei.com}

\renewcommand{\shortauthors}{Anon.}

\begin{abstract}
This work proposes to analyse some keywords for bias analysis. For this, we are using several NLP approaches and compare them based on their capability of analysing keywords to analyse bias.
The overall findings show that our proposed approach gives comparable results with state-of-the-art.
\end{abstract}

%
%


\maketitle

\section{Introduction}
\label{sec:introduction}

Key-phrase extraction aims to choose a set of important words or phrases automatically that summarize the main points of a given document. Key-phrases can help readers to search information quickly and accurately, especially in large text collections.
Researchers have already proposed various approaches to extract key-phrases. For an in-depth study, please refer to~\cite{hasan2014automatic}.

The problem of extracting key-phrases can be mainly solved using two different kinds of approaches: supervised and unsupervised. Although supervised approaches generally perform better on domain-specific tasks, data annotation is a costly process, and the trained models might not be leveraged for other datasets. Unsupervised models can easily be applied to distinct datasets with no customization or little modification needed, i.e., one can only use the suitable models for different languages, yet they can provide comparable results.

Successful unsupervised key-phrase extraction methods are composed of graph-based and embedding-based approaches~\cite{bennani2018simple, boudin2018unsupervised, bougouin2013topicrank, sun2020sifrank, wan2008collabrank}.
SingleRank~\cite{wan2008collabrank} initially constructs a set of neighbour documents for each document $d$.
Then, this set is used to generate candidate key-phrases with a graph ranking algorithm that has both local (of $d$ itself) and global (of the whole neighbor document set) components. Unlike SingleRank which also uses corpus information that a given document might be linked to, sBERT-Rank only requires the given document at a time. From the given document, sBERT-Rank first detects the noun-phrases with the Stanza dependency parser
Compared to SingleRank~\cite{wan2008collabrank} this is a more efficient and realistic approach. In this sense, sBERT-Rank is a corpus-independent approach that leverages the language information embedded in the pre-trained model of BERT~\cite{devlin2018bert}.
On the other hand, TopicRank~\cite{bougouin2013topicrank} constructs a graph of clusters of candidate phrases for various topics regarding the given document. Similar to SingleRank~\cite{wan2008collabrank}, a graph ranking algorithm is utilized to choose the top clusters (nodes). Finally, the key-phrases are extracted from these top topics. In a follow-up work, namely MultipartiteRank~\cite{boudin2018unsupervised}, the authors additionally propose to add candidate phrases as nodes to the graph in a multipartite graph structure. Nonetheless, the major difference from TopicRank~\cite{bougouin2013topicrank} is that MultipartiteRank~\cite{boudin2018unsupervised} adds position information to the edge weights to capture the semantics of longer documents.

In addition to those graph-based methods, 
embedding-based approaches have also been used for extracting key-phrases, since external information from pre-trained language models can help to improve the model capability~\cite{bennani2018simple, sun2020sifrank}.
Our proposed pipeline is similar to the embedding-based ranking methods of EmbedRank~\cite{bennani2018simple} and SIFRank~\cite{sun2020sifrank}. EmbedRank generates candidate phrases by initially extracting all possible phrases between the lengths of 2 to 8 tokens from a given document by only using POSTag information
Then, these phrases, along with the original document, are embedded into a vector space using the embedding methods of Sent2Vec~\cite{pagliardini2017unsupervised} and Doc2Vec~\cite{lau2016empirical}. Further, these phrases are ranked with respect to their cosine similarities with the corresponding document from which they were extracted. 
sBERT-Rank is mainly different from EmbedRank in two ways. sBERT-Rank uses a dependency tree instead of only POSTag information to generate candidate phrases (noun-phrases) in a more efficient and effective manner. In this way, our approach firstly selects suitable key-phrases using word relations rather than applying simple word permutations, which brings extra computation for less informative phrases. The second difference is that our key-phrase extraction pipeline uses sBERT~\cite{reimers2019sentence} which also leverages the hidden positional embedding mechanism of BERT instead of the embedding methods of Sent2Vec and Doc2Vec.
%


In addition, in SIFRank~\cite{sun2020sifrank}, the authors suggest to firstly extract the topic of a given document by using sentence embeddings. Then, the candidate phrases are compared with the topic (not the sentences or the document itself) for ranking. Subsequently, a position-biased weight is assigned to each candidate phrase to improve the model's performance on long documents. 
Unlike the SIFRank, the performance of the sBERT-Rank is not affected by the document length at all since the ranking score is computed in two steps. 
These subsequent steps help us to select the candidate phrases that are informative of the corresponding sentence from which the candidates were extracted, and the sentence is informative of the given document. Apart from this, SIFRank uses~\cite{arora2017simple}, while we use sBERT for sentence embeddings.

%
In this work, we introduce a simple, corpus-independent, i.e., only requires the current document itself, and an easily-extendible to other languages, unsupervised embedding-based pipeline to extract key-phrases from a given document. To the best of our knowledge, this is the only work that uses all four diverse benchmark datasets in the literature to properly evaluate the capability of our proposed pipeline of sBERT-Rank as well as provide a comprehensive evaluation of other state-of-the-art approaches. This is important in the sense that unsupervised approaches are generally expected to give comparable and consistent results irrespective of the dataset properties, i.e. longer or shorter documents, news articles of scientific papers etc., thus we believe that it is also necessary to assess this property of our pipeline as well as the-state-of-the-art approaches.

The remainder of the paper is structured as follows. In Section~\ref{sec:our_method} we present the details of our pipeline, sBERT-Rank. In Section~\ref{sec:experiments} we provide the experimental results and then discuss them. About the experimental setup and then display the results. Lastly, we conclude the paper in Section~\ref{sec:conclusion}.

\section{\makefirstuc{sBERT-Rank}}
\label{sec:our_method}

\begin{figure}[!h]
\vspace{-1.6em}
\centering
\begin{tikzpicture} [
block/.style={rectangle, draw=black!60, very thick, minimum width=30mm,  minimum height=8mm, node distance=0.45cm},
] 

\node[] (input) {{sent}};

\node[block] (np_chunker) [below=of input] {Noun-Phrase Chunker};

\node[] (output_1) [below=of np_chunker] {list of ${NP}$};

\node[block] (sbert) [below=of output_1] {sBERT};

\node[] (output_2) [below=of sbert] {list of candidate embeddings};

\node[block] (rank_cand) [below=of output_2] {Ranking Candidates};

\node[] (output_3) [below=of rank_cand] {informativeness of candidates};

\draw[->] (input.south) -- (np_chunker.north);
\draw[->] (np_chunker.south) -- (output_1.north);
\draw[->] (output_1.south) -- (sbert.north);
\draw[->] (sbert.south) -- (output_2.north);
\draw[->] (output_2.south) -- (rank_cand.north);
\draw[->] (rank_cand.south) --(output_3.north);

\end{tikzpicture}
\vspace{-1em}
\caption{\label{fig:sBERT_figure}sBert-Rank Pipeline}
\vspace{-1.2em}
\end{figure}
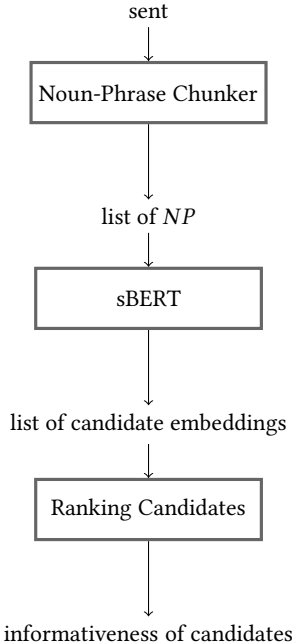

In this section, we present the core components of our unsupervised key-phrase extraction pipeline. Our pipeline is composed of three main parts. First, we extract informative noun-phrases as candidates from a given document. Then, these candidates and the current document are represented in a high dimensional vector space using sBERT. Lastly, we rank the candidate phrases by using the cosine similarity scores. A simple flowchart explaining this process can be seen in \ref{fig:sBERT_figure}.

\subsection{Noun-Phrase Chunker}
\label{sec:nounphrase}
Key-phrases are the most informative keywords or 
multiple words for the document at hand. They are mostly nouns as well as noun or adjective phrases. Hence, in this work, we aim to catch noun-phrases including adjective phrases within a given document.

In order to detect noun-phrases, a widely-used stable dependency parser of Stanza~\cite{qi2020stanza} proposed by the Stanford NLP Group is used. The current document is first split into , and these sentences are given to the Stanza as inputs to generate its dependency tree for the sentences in the document. This allows us to generate dependency trees both for long and short documents. Since dependency trees contain valuable word- information, after creating the dependency tree for a given document, its noun-phrases are extracted with a rule-based approach which leverages the dependency tree information.

To exploit dependency tree information, we propose a one general rule where we consider a noun-phrase (\emph{NP}) as a recurrence relation. The general rule is defined as follows:

\begin{linenomath*}
\vspace{-0.4em}
\begin{equation}\label{eq:delta_bias}
    NP = (NOUN, ADJ)(NP)
\end{equation}
\vspace{-0.4em}
\end{linenomath*}

In the given rule, $NP$ initially starts with a noun or adjective, then it is followed by a noun-phrase itself. Furthermore, if we consider a given ${NP}_i$, which is comprised of $n$ tokens in total we define each of these tokens as $t_j$, where $j = 1, 2, 3, ... n$. Note that ${NP}_i$ is defined as a list of $n$ tokens with the order of ${t_1, t_2, t_3, ..., t_n}$. Then, our rule requires that the head of each $t_i$, $ 1 \leq i \leq n$, extracted from the dependency tree, which is denoted as $h(t_i)$, should be $t_n$. That is, all of the tokens within the noun-phrase should be modified by the last token. 
This condition helps us to filter the noun-phrases in such a way that each candidate is a complete phrase by itself. Thus, as mentioned in Section~\ref{sec:introduction} differently from~\cite{bennani2018simple} which uses only POSTag information in the candidate generation step, we obtain a more distilled set of candidate phrases using dependency tree information. In this way, the distilled set generally provides more informative candidates which are smaller in size than the ones generated by~\cite{bennani2018simple} which could provide a more effective as well an efficient processing of the documents in a given dataset. Note that for each document, only distinct noun-phrases are taken. Apart from these, since stanza is available for more than 70 languages~\footnote{\url{https://stanfordnlp.github.io/stanza/}}, the first step of our pipeline can easily be extended to extract key-phrases in other languages.

\begin{table*}[!t]
    \vspace{-0.8em}
    \centering
    \caption{Performance of sBERT-Rank on INSPEC, DUC2001, NUS and SemEval2017 Datasets}
    \vspace{-1em}
    \setlength{\tabcolsep}{3.0pt}
    \renewcommand{\arraystretch}{1.0}
    \begin{tabular}{c|c|ccc|ccc|ccc|ccc}
        \hline\hline
        & & &  INSPEC & & &  DUC2001 & & &  NUS & & & SemEval2017\\
        \hline
        N & Model & R & P & F1 & R & P & F1 & R & P & F1 & R & P & F1 \\
        \hline\hline 
        \multirow{6}{*}{5} 
        & MultipartiteRank &
        0.21 & 0.28 & 0.24 & 
        0.18 & 0.15 & 0.16 &
        0.10 & 0.19 & 0.13 &
        0.11 & 0.34 & 0.16\\
        & SingleRank &
        0.2 & 0.26 & 0.23 &
        0.11 & 0.10 & 0.10 &
        0.02 & 0.04 & 0.03 &
        0.11 & 0.34 & 0.17 \\
        & TopicRank &
        0.21 & 0.28 & 0.24 &
        0.16 & 0.14 & 0.15 &
        0.09 & 0.17 & 0.12 & 
        0.10 & 0.34 & 0.16 \\
        & EmbedRank &
        0.25& 0.41& 0.31&
        0.22 & 0.34 & 0.27&
        0.02 & 0.05 & 0.03&
        0.13 & 0.45 & 0.20 \\
        & SIFRank &
        0.21 & \textbf{0.43} & \textbf{0.29} &
        \textbf{0.25} & \textbf{0.40} & \textbf{0.31} &
        0.03 & 0.04 & 0.03 & 
        \textbf{0.15} & \textbf{0.50} & \textbf{0.23}\\
        & sBERT-Rank &
        \textbf{0.28} & 0.29 & \textbf{0.29} &
        0.24 & 0.22 & 0.23 &
        \textbf{0.14} & \textbf{0.18} & \textbf{0.15} &
        0.13 & 0.32 & 0.19 \\
        
        \hline
        \multirow{6}{*}{10} 
        & MultipartiteRank &
        0.33 & 0.24 & 0.28 &
        0.26 & 0.12 & 0.17 &
        0.17 & \textbf{0.16} & \textbf{0.17} & 
        0.18 & 0.28 & 0.22\\
        & SingleRank &
        0.37 & 0.26 & 0.30 &
        0.22 & 0.11 & 0.14 &
        0.05 & 0.05 & 0.05 & 
        0.21 & 0.33 & 0.26\\
        & TopicRank &
        0.31 & 0.23 & 0.27 &
        0.25 & 0.11 & 0.15 &
        0.14 & 0.14 & 0.14 & 
        0.17 & 0.27 & 0.21 \\
        & EmbedRank &
        0.40 & 0.35  & 0.37 &
        0.35 &  0.28 & 0.28 &
        0.05 & 0.06 & 0.05 & 
        0.23 & 0.40 & 0.30 \\
        & SIFRank &
        0.39 & \textbf{0.39} & \textbf{0.39} &
        0.37 & \textbf{0.30} & \textbf{0.33} &
        0.04 & 0.04 & 0.04 & 
        \textbf{0.26} & \textbf{0.45} & \textbf{0.33} \\
        & sBERT-Rank &
        \textbf{0.43} & 0.24 & 0.31 &
        \textbf{0.39} & 0.19 & 0.25 &
        \textbf{0.22} & 0.14 & \textbf{0.17} & 
        0.24 & 0.29 & 0.26\\
        
        \hline
        \multirow{6}{*}{15} 
        & MultipartiteRank &
        0.43 & 0.22 & 0.29&
        0.33 & 0.11 & 0.16 &
        0.22 & \textbf{0.14} & \textbf{0.17} & 
        0.24  & 0.26  & 0.25 \\
        & SingleRank &
        0.47 & 0.24 & 0.32 &
        0.29 & 0.09 & 0.15 &
        0.08 & 0.05 & 0.06 &
        0.29 & 0.30 & 0.29 \\
        & TopicRank &
        0.38 & 0.21 & 0.27 &
        0.31 & 0.09 & 0.14 &
        0.17 & 0.11 & 0.13 & 
        0.22 & 0.23 & 0.22 \\
        & EmbedRank &
        0.49 & 0.31 & 0.38 &
        0.44 & 0.24 &  0.31 & 
        0.07 & 0.05 & 0.06 & 
        0.32 & 0.37 & 0.34 \\
        & SIFRank &
        0.48 & \textbf{0.33} & \textbf{0.39} &
        0.45 & \textbf{0.25} & \textbf{0.32} &
        0.06 & 0.04 & 0.05 & 
        \textbf{0.35} & \textbf{0.41} & \textbf{0.38} \\
        & sBERT-Rank &
        \textbf{0.52} & 0.21 & 0.30 &
        \textbf{0.50} & 0.16 & 0.24 &
        \textbf{0.28} & 0.12 & \textbf{0.17} & 
        0.32 & 0.26 & 0.29 \\
        
        \hline\hline
    \end{tabular}
    \label{tab:key-phrase}
    \vspace{-1em}
\end{table*}

\subsection{sBERT}
After obtaining the unique noun-phrases which are the candidates for each document, it is necessary to assess the informativeness level of each unique noun-phrase for the document to which it belongs. For this, candidate phrases and the document itself are represented in a high-dimensional vector space by using sBERT embeddings. sBERT modifies the pre-trained BERT network by using siamese networks to derive semantically meaningful sentences. Yet, in the scope of this work, we use sBERT not only for sentences, but for computing semantic textual similarity~\footnote{\url{https://www.sbert.net/}}. sBERT specifically provides a pre-trained model for English as well as a multilingual model for other languages. Thus, in addition to the first step of sBERT-Rank, the second step can easily be extended to other languages as well by using the multilingual model of sBERT.



\subsection{Ranking Candidates}
In the last step of our pipeline, for ranking the candidate phrases that are represented in a high-dimensional space, two cosine similarity scores are computed. First, the cosine similarity scores between each sentence in the current document and the document itself are computed. Then, the cosine similarity scores between each candidate phrase and the sentence from which the phrase is extracted are also computed. In this way, the informativeness level of each candidate phrase is calculated by multiplying the computed cosine similarity scores. After that, the candidate phrases are ranked with respect to their informativeness level scores in ascending order. 

Let's assume that a given document ${doc}_i$ contains $n$ sentences. As the first step, ${doc}_i$ is splitted into its sentences of ${sent}_j$ for $j = 1, 2, 3, ... n$. Then, each sentence ${sent}_j$ is given to the Stanza and based on the dependency tree information and the rule-based approach let's assume that in total $m$ candidate phrases are extracted from ${sent}_j$. In this case, for each candidate phrase ${cand}_k$ where $k = 1, 2, 3, ... m$, the informativeness level of ${inf}_{{cand}_k}$ is computed as follows:

\begin{equation}
    \label{eq:similarity}
    {inf}_{{cand}_k} = cos ({cand}_k, {sent}_j) * cos ({sent}_j, {doc}_i)
    \vspace{1em}
\end{equation}

Note that $cos(a,b)$ computes the cosine similarity score between the textual contents of $a$ and $b$. In Eq.~\eqref{eq:similarity} the first component is the cosine similarity score computed between the candidate phrase and the sentence; while, the second component denotes the similarity score between the sentence and the whole document. After computing the score of ${inf}_{{cand}_k}$ for all $m$ extracted candidate phrases in ${sent}_j$, we fulfill the same computation for all the candidate phrases of $n$ sentences. Then, for each document ${doc}_i$, all of these scores are aggregated into a list which is sorted with respect to the computed informativeness level scores -- the candidate phrases with higher scores are expected to be more informative about ${doc}_i$.





\section{Experimental Setup}
\label{sec:experiments}

%
In this section, we give the details of our experiments. For this, first we provide information about the evaluation datasets, and then we report the results as displayed in Table~\ref{tab:keyphrase} and discuss them.

\subsection{Dataset}
\label{sec:dataset}
For evaluating our key-phrase pipeline capability, we used the same four benchmark datasets, namely~\emph{INSPEC},~\emph{DUC},~\emph{NUS}, and~\emph{SemEval2017} that have been used in unsupervised key-phrase extraction literature.
~\emph{INSPEC} is a scientific paper abstract dataset that contains around $2000$ abstracts with the assigned key-phrases.
On the other hand,~\emph{DUC} is composed of $308$ newspaper articles and the articles are annotated with key-phrases from TREC-9 by the articles' authors.~\emph{NUS} is a dataset of $211$ full scientific papers, and the papers are assigned with key-phrases by both authors and annotators.
Lastly,~\emph{SemEval2017} is a dataset of $493$ paragraphs selected from ScienceDirect~\footnote{\url{https://www.sciencedirect.com/}} articles, in the research fields of Computer Science, Material Sciences, and Physics. The articles are assigned with key-phrases by undergraduate students and expert annotators.

Based on the information, datasets show distinct properties since they contain various kinds of textual content, i.e. news articles, scientific papers, etc., as well as their annotation procedure for assigning the key-phrases are different from each other.
Apart from these, based on the dataset statistics in Table~\ref{tab:datasets}, the datasets are also dissimilar in the sense that they contain longer or shorter documents. For instance, ~\emph{INSPEC} is composed of shorter documents while~\emph{NUS} includes x40 longer documents. Additionally, datasets include a different number of key-phrases on average for each document. All of these diverse dataset properties can affect the key-phrase extraction capability of a given model; therefore we believe that our evaluation results are sufficiently comprehensive. Nonetheless, the-state-of-the-art approaches displayed in  Table~\ref{tab:key-phrase} evaluated their models only on three benchmark datasets, MultipartiteRank and EmbedRank used~\emph{INSPEC},~\emph{DUC},~\emph{NUS} whereas SIFRank used~\emph{INSPEC},~\emph{DUC}, and~\emph{SemEval2017}. Finally, SingleRank used only~\emph{DUC} and TopicRank utilized~\emph{INSPEC}, ~\emph{SemEval2010}, ~\emph{WikiNews} and~\emph{DEFT}.



\begin{table}[!h]
    \vspace{-0.5em}
    \centering
    \begin{tabular}{c|ccccc}
         & INSPEC & DUC2001 & NUS & SemEval2017\\
         \hline
         \# key-phrases & 14.11 & 8.07 & 11.71 & 17.30 \\
         \# tokens & 124.25 & 732.78 & 7056.17 & 168.86 \\
         \# documents & 2000 & 308 & 211 & 493 \\
    \end{tabular}
    \caption{Average number of tokens and key-phrases for each document in the four benchmark datasets used.}
    \label{tab:datasets}
   \vspace{-3em}
\end{table}

\subsection{Results \& Discussion}
\label{sec:results}
Our results in Table~\ref{tab:key-phrase} show that sBERT-Rank outperforms the state-of-the-art unsupervised key-phrase extraction approaches in the~\emph{NUS} dataset in terms of recall values, and it also gives comparable F1-scores with the other methods.
Moreover, sBERT-Rank achieves the highest recall values on the~\emph{INSPEC} and~\emph{DUC} datasets and comparable results in terms of F1-scores, especially on the~\emph{INSPEC} dataset. Apart from this, sBERT-Rank achieves the second-highest recall values on the~\emph{SemEval2017} dataset after the SIFRank, and gives the third-highest F1-scores among the other approaches. Even if the SIFRank surpasses sBERT-Rank on the~\emph{SemEval2017}, it gives very low scores of all precision, recall and F1-scores on the~\emph{NUS} dataset -- in terms of F1-scores for instance sBERT-Rank outweighs the SIFRank by at least 3.5 times. This shows that the SIFRank cannot provide consistent results, probably for datasets containing very long documents like the~\emph{NUS} dataset. It might not be successful for extracting key-phrases. Unlike our proposed pipeline, sBERT-Rank gives consistent results irrespective of the dataset properties. This is the only pipeline in Table~\ref{tab:key-phrase} that either outperforms other approaches, or achieves comparable scores (among the top-3 highest scores). We believe that achieving comparable results on different datasets regardless of their instrinsic properties is an important property for an unsupervised pipeline that is expected to be leveraged on diverse datasets from various domains in practice.

This important property of our introduced pipeline sBERT-Rank can probably be attributed to the fact that the informativeness level of a given candidate phrase is computed by multiplying the two similarity scores. The candidate can obtain a high informativeness score, only if it is informative of the corresponding sentence as well as if the sentence is informative of the document at hand. This step not only helps us to select the most informative candidate phrases from the first set of extracted ones, but it also alleviates the scalability-related issues that most of the aforementioned state-of-the-art approaches possibly suffer in practice. Note that regarding the scalability, not only the second step of selecting informative candidates based on their informativeness scores, but also the initial step of extracting the first set of candidate phrases helps since we obtain a smaller first set since we have an additional constraint that was mentioned in Section~\ref{sec:nounphrase}.






In addition to these, our proposed pipeline sBERT-Rank has its own limitations. Except for the~\emph{NUS} dataset, sBERT-Rank gives comparable yet lower precision scores. Improving the precision scores by examining the experimental results with the corresponding dataset properties is left as future work. Nonetheless, the results indicate that the precision scores of sBERT-Rank gradually decrease while their recall scores gradually increase when the candidate number increases from $5$ to $15$ on all the four benchmark datasets. For reproducibility purposes, you can find our code at~\repourl.

The hypothetical function generating the key-phrases might be way to complex to be approximated by the aforementioned scores. Thus, as the number of candidates increase, the approximation that supposedly works with a small number of candidates might diverge from being a good predictor. Which is counter-intuitive to the fact that recall values increase in all of the methods as the number of candidates increases but is also reasonable when we consider that the number of key-phrases to detect is the same and generated by a constant hypothetical base distribution.  


\section{Conclusion \& Future Work}
\label{sec:conclusion}
In this work, we proposed a new unsupervised pipeline, sBERT-Rank which is a simple, easily-extendible and corpus-independent pipeline for key-phrase extraction. sBERT-Rank leverages noun-phrases extracted from dependency parsing in order to detect candidate phrases with the rule-based approach. This step helps us to create a good set of informative candidate phrases. Then, using sBERT, candidate phrases, sentences, and documents are all represented in a high-dimensional vector space, which serves to capture the semantic similarity of the given textual contents.
The scoring method of sBERT-Rank provides the user with informativeness scores for each candidate regardless of its position in the given text. Furthermore, the scoring mechanism with two components does not favor a candidate if its sentence not relevant or not informative for the current document at hand.



As a future work, we aim to evaluate the model capability of sBERT-Rank on other languages since it is easily-extendible. Nonetheless, to improve the overall precision scores and to alleviate the problem of gradually decreasing precision scores, different scoring functions can be investigated. Moreover, key-phrase results with the corresponding dataset properies can be elaborately examined for finding better scoring functions or the pipeline steps can be slightly modified to prevent the gradual increase/decrease of recall/precision scores when the candidate number increases.


\newpage

\bibliographystyle{ACM-Reference-Format}
\bibliography{main}

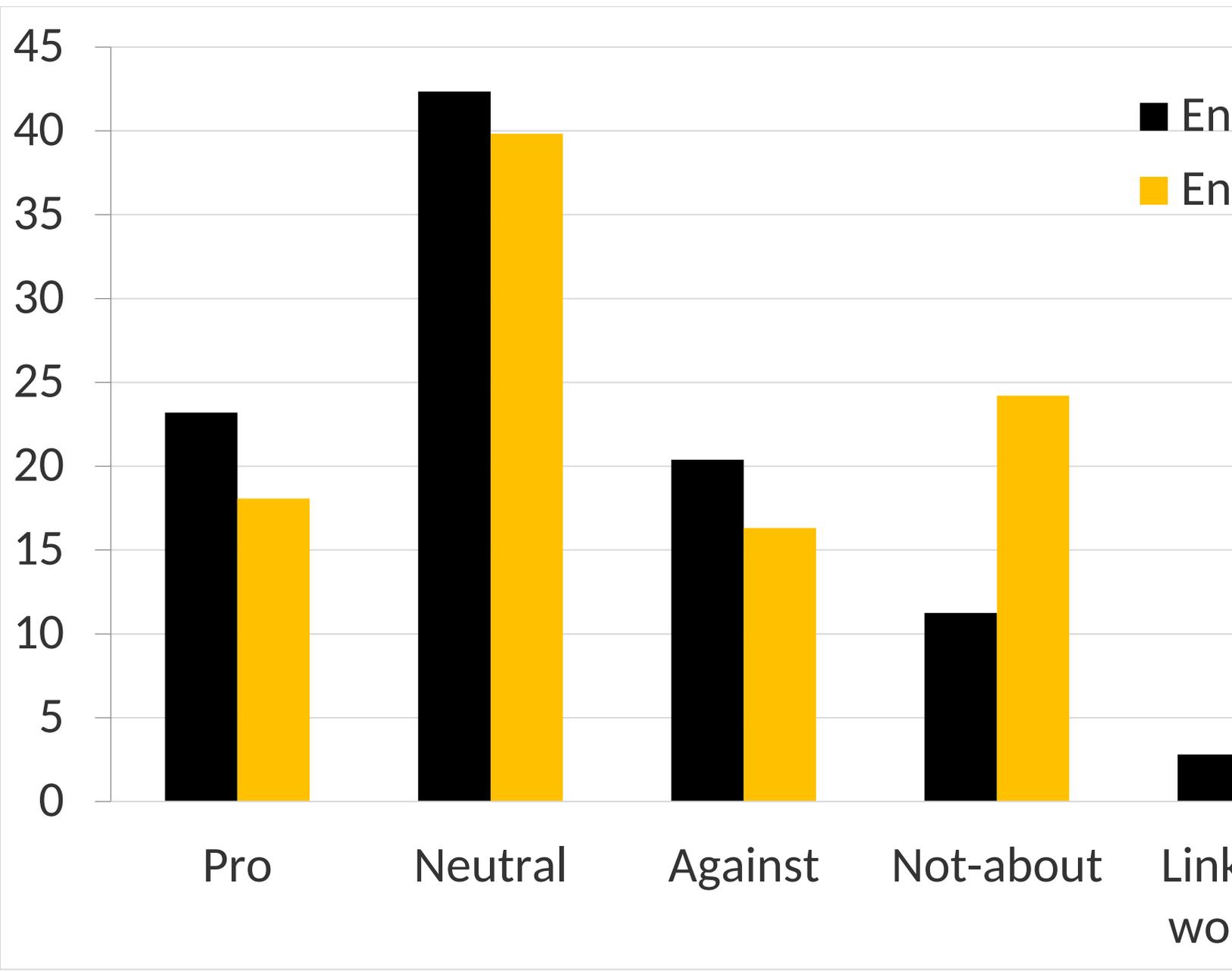
\end{document}